\documentclass[%
 aip,
 amsmath,amssymb,
 reprint,%
]{revtex4-1}

\usepackage{graphicx}
\usepackage{dcolumn}
\usepackage{bm}

\usepackage[utf8]{inputenc}
\usepackage[T1]{fontenc}
\usepackage{mathptmx}
\usepackage{amsmath}
\usepackage{accents}
\usepackage{subfig}
\DeclareMathOperator{\Tr}{tr}
\usepackage{color}
\usepackage[linktocpage,colorlinks=true,linkcolor=blue,citecolor=blue,breaklinks=true]{hyperref}
\usepackage{verbatim}

\usepackage{ulem}

\renewcommand{\vec}[1]{\underline{#1}}
\newcommand{\tens}[1]{\underline{\underline{#1}}}
\newcommand{\bnabla}{\vec{\nabla}}

\begin{document}

\preprint{AIP/123-QED}

\title{Topological states in chiral active matter: dynamic blue phases and active half-skyrmions}%

\author{Luuk Metselaar}
\affiliation{The Rudolf Peierls Centre for Theoretical Physics, Clarendon Laboratory, Parks Road, Oxford OX1 3PU, UK}
\author{Amin Doostmohammadi}
\email{amin.doostmohammadi@physics.ox.ac.uk}
\affiliation{The Rudolf Peierls Centre for Theoretical Physics, Clarendon Laboratory, Parks Road, Oxford OX1 3PU, UK}
\author{Julia M. Yeomans}
\email{julia.yeomans@physics.ox.ac.uk}
\affiliation{The Rudolf Peierls Centre for Theoretical Physics, Clarendon Laboratory, Parks Road, Oxford OX1 3PU, UK}

\date{\today}

\begin{abstract}
We numerically study the dynamics of two-dimensional blue phases in active chiral liquid crystals. We show that introducing contractile activity results in stabilised blue phases, while small extensile activity generates ordered but dynamic blue phases characterised by coherently moving half-skyrmions and disclinations. Increasing extensile activity above a threshold leads to the dissociation of the half-skyrmions and active turbulence. We further analyse isolated active half-skyrmions in an isotropic background and compare the activity-induced velocity fields in simulations to an analytical prediction of the flow. Finally, we show that confining an active blue phase can give rise to a system-wide circulation, in which half-skyrmions and disclinations rotate together. 
\end{abstract}

\maketitle

\section{Introduction}
Active matter is continuously driven out of equilibrium by the injection of energy at the level of its individual constituent particles, leading to phenomena such as spontaneous flow generation \cite{Simha2002,Voituriez2005} and active turbulence \cite{Wensink2012,Giomi2015}. Furthermore, liquid crystal features including long-range orientational order and topological defects emerge in systems composed of anisotropic active particles, such as rod-shaped bacteria~\cite{Volfson08,Nishiguchi17} or cytoskeletal filaments~\cite{Sanchez12}. Owing to their widespread relevance to biological systems - from subcellular actomyosin mixtures~\cite{Kumar18} to bacterial biofilms~\cite{You18} and tissues~\cite{Gruler99,Saw17,Kawaguchi17,Duclos17} - active liquid crystals have become one of the key model systems for understanding active materials~\cite{Prost2015,Doost18}.

Previous studies of active liquid crystals have however mostly focused on achiral particles. In nature, chirality is ubiquitous, ranging from the helical structure of DNA \cite{Watson1953} and bacterial flagella \cite{Berg1973} to numerous biopolymers such as actin, chitin and microtubules \cite{Bouligand1978}. On a more macroscopic level, the mitotic spindle is chiral due to torques within microtubule bundles \cite{Novak2018}, cells can develop chiral actomyosin patterns \cite{Tee2015}, and there is even tissue-scale chirality in spontaneous cellular shear flow \cite{Duclos2018}. Motivated by this common occurrence of chirality in biological systems, in this paper we explore the combined effects of activity and chirality on pattern formation within active chiral liquid crystals.

In the absence of any activity, a particular case of chiral patterning in liquid crystals occurs when there is helical ordering in more than one direction. Topological constraints mean that it is not possible to construct such a state without introducing defects or disclination lines. If the free energy advantage of twisting offsets the cost of introducing topological defects it is however possible to obtain local regions of double-twist surrounded by negative disclinations positioned, in two dimensions, in a hexagonal lattice. This structure, occurring in the temperature range between the isotropic and the cholesteric phases, is generally known as a blue phase in the context of passive liquid crystals~\cite{Wright1989}.

The double-twisted regions can be regarded as vortex-like excitations without singularities at their centre, referred to as `half-skyrmions' (Fig. \ref{fig:schematic}). Although the original work by Skyrme \cite{Skyrme1962} dealt with three-dimensional topological excitations, two-dimensional full- and half-skyrmions have recently attracted considerable attention in condensed matter systems \cite{Ackerman2015,Jiang2015,Guo2016,Yu18}. A lattice of two-dimensional half-skyrmions has been observed experimentally in the precursor state of a bulk cubic helimagnet using neutron scattering \cite{Pappas2009}, in atomically thin ferromagnets films with scanning tunnelling microscopy  \cite{Heinze2011}, and recently also in thin films of passive chiral liquid crystals by direct optical means \cite{Nych2017}.

In this paper we simulate quasi-two-dimensional active chiral liquid crystals. We show that stable vortex lattices form due to the presence of internal chiral structure in an active two-dimensional blue phase, and demonstrate a threshold for dissociation into unbound vortices for the case of extensile activity. For contractile activity, however, the vortex lattice remains stable for any activity strength considered. Furthermore, we compare analytical and computational flow-fields around isolated active half-skyrmions in an isotropic background, and show that these structures are stable under contractile activity, but unstable for the case of extensile activity. This is consistent with the suppression of the splay instability reported for active cholesterics \cite{Whitfield2017}, but can more intuitively be understood in terms of the stability of the half-skyrmions. Finally, we simulate an active two-dimensional blue phase in square confinement and find a dynamically ordered state where the blue phase rotates collectively as a coherent macroscopic unit.

\section{Governing Equations}
In a recent paper, we simulated quasi-2D membranes of a passive cholesteric liquid crystal using a nematohydrodynamic approach~\cite{Metselaar2018}. Here, we use the nematohydrodynamic equations of liquid crystals, modified to account for the stresses generated by active constituent elements \cite{Simha2002,Whitfield2017}. The total density $\rho$ and the velocity field $\vec{u}$ obey the incompressible Navier-Stokes equations
\begin{equation}
\bnabla \cdot \vec{u} = 0,
\label{eq:mass}
\end{equation}
\begin{equation}
\rho \left(\partial_t + \vec{u} \cdot \bnabla\right)\vec{u} = \bnabla \cdot \tens{\Pi}
\label{eq:momentum}
\end{equation}
where $\tens{\Pi}$ is the stress tensor.
To account for the macroscopic orientational order of the microscopic active and anisotropic particles, the nematic tensor $\tens{Q} = \frac{3S_0}{2}\left(\vec{n}\vec{n}-\tens{I}/3\right)$ is introduced. This tensor is symmetric and traceless, and $S_0$ is the coarse-grained magnitude of the nematic order, $\vec{n}$ is the local nematic director and $\tens{I}$ is the identity matrix. The nematic tensor evolves as~\cite{BerisBook}
\begin{equation}
\left(\partial_t + \vec{u} \cdot \bnabla\right)\tens{Q} - \tens{S} = \Gamma\tens{H}
\label{eq:orientation}
\end{equation}
with $\Gamma$ the rotational diffusivity. The co-rotation term $\tens{S}$, given by
\begin{equation}
\begin{split}
\tens{S} = \left(\xi\tens{E} + \tens{\Omega}\right)\cdot\left(\tens{Q} + \tens{I}/3\right) + \\ \left(\tens{Q} + \tens{I}/3\right)\cdot \left(\xi\tens{E} - \tens{\Omega}\right) - 2\xi\left(\tens{Q} + \tens{I}/3\right)\left(\tens{Q}:\bnabla\vec{u}\right),
\end{split}
\end{equation}
accounts for the response of the orientational order to flow gradients described by the strain rate tensor $\tens{E} = \left(\bnabla\vec{u}^T+\bnabla\vec{u}\right)/2$ and the vorticity tensor $\tens{\Omega} = \left(\bnabla\vec{u}^T-\bnabla\vec{u}\right)/2$. The alignment parameter $\xi$ characterises the response of the director to the strain rate and vorticity tensors and depends on particle shapes: $\xi >0$, $\xi < 0$ and $\xi = 0$ correspond to rod-like, disk-like, and spherical particles, respectively. The relaxation of the orientational order is governed by the molecular field 
\begin{equation}
\tens{H} = -\frac{\delta \mathcal{F}}{\delta \tens{Q}} + \frac{\tens{I}}{3}\Tr{\frac{\delta \mathcal{F}}{\delta \tens{Q}}},
\end{equation}
where $\mathcal{F} = \mathcal{F}_\text{b} + \mathcal{F}_\text{el}$ is the free energy composed of the bulk term~\cite{DeGennes1995}
\begin{equation}
\mathcal{F}_\text{b} = A_0\left(\frac{1}{2}\big(1-\frac{\gamma}{3}\big)\tens{Q}^2 - \frac{\gamma}{3}\tens{Q}^3 + \frac{\gamma}{4}\tens{Q}^4\right),\label{eq:LDG}
\end{equation}
and the elastic contribution 
\begin{equation}
\mathcal{F}_\text{el} = \frac{1}{2}L_1\left(\bnabla \cdot \tens{Q}\right)^2 + \frac{1}{2}L_C\left(\bnabla \times \tens{Q} + 2q_0\tens{Q}\right)^2,\label{eq:Frank}
\end{equation}
where $A_0$, $\gamma$, $L_1$, $L_C$, and $q_0$ are material constants. In particular, the Landau-de Gennes bulk free energy (Eq.~\ref{eq:LDG}) describes a first order isotropic-nematic phase transition at $\gamma = 2.7$. For $\gamma < 2.7$ the liquid crystal will be in the isotropic phase, while it will be in the nematic or cholesteric phase for $\gamma > 2.7$. The Frank free energy (Eq.~\ref{eq:Frank}) penalises orientational deformations, with $L_1$ and $L_C$  denoting the elastic constants and $q_0$ setting the inverse pitch for a cholesteric liquid crystal. 

In addition to the usual viscous stress $\tens{\Pi}_\text{visc} = 2\eta\tens{E}$, the stress term in equation \ref{eq:momentum} accounts for contributions from the elastic stresses and the activity. The elastic stresses, generating backflow, are given by~\cite{BerisBook}
\begin{equation}
\begin{split}
\tens{\Pi}_\text{el} = -P\tens{I} + 2\xi\left(\tens{Q}+\tens{I}/3\right)\left(\tens{Q}:\tens{H}\right) - \\ \xi\tens{H} \cdot \left(\tens{Q}+\tens{I}/3\right) - \xi \left(\tens{Q}+\tens{I}/3\right) \cdot \tens{H} - \\ \bnabla \tens{Q} : \frac{\delta \mathcal{F}}{\delta(\bnabla\tens{Q})} + \tens{Q} \cdot \tens{H} - \tens{H} \cdot \tens{Q},
\end{split}
\end{equation}
which includes the pressure, $P$. The active stress is proportional to the nematic tensor $\tens{\Pi}_\text{act} = -\zeta\tens{Q}$, such that any gradient in the nematic tensor generates a flow field, with strength determined by the activity parameter, $\zeta$. Positive $\zeta$ corresponds to extensile activity, where constituent active particles drag fluid in towards their sides and expel it along their elongation axis. For contractile activity $\zeta<0$, and fluid is pulled in along the length of the particles and pushed out from their sides. An explicit chiral active term such as the one introduced in~\cite{Maitra2018} is not necessary in our case since we are dealing with the full three-dimensional nematic tensor, while the two-dimensional tensor in~\cite{Maitra2018} does not include any twist.

\section{Results}
\subsection{Dynamic blue phases}
\begin{figure}
\centering
\includegraphics[width=0.5\linewidth]{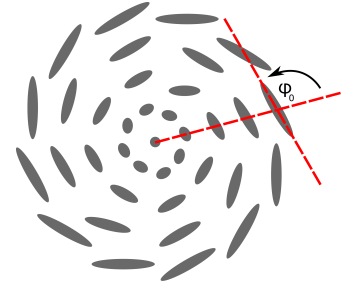}
\caption{Top view of a single half-skyrmion. The director twists through an angle of $\pi/2$, from a vertical orientation in the centre of the half-skyrmion to a horizontal orientation at the edge. The angle $\phi_0$ is $\pi/2$ in equilibrium, but changes under influence of activity, leading to a `swirl'-like profile.}
\label{fig:schematic}
\end{figure}

The active nematohydrodynamics equations \ref{eq:mass}, \ref{eq:momentum} and \ref{eq:orientation} are solved using a hybrid lattice Boltzmann - finite difference method \cite{Marenduzzo2007,Thampi2014}. The time step and lattice spacing are set to unity. The parameters used are $A_0 = 1.5$, $\gamma = 2.85$, $L_1 = 0.04$, $L_C = 0.085$, $q_0 = 2\pi/10.3$, $\Gamma = 0.7$, $\xi = 0.5 - 0.9$, $\eta = 2/3$, $\rho = 1$, in lattice Boltzmann units. We model a two-dimensional velocity field, but in order to allow for twist the director is  free to point in three dimensions. For sufficiently high chirality, the ground state of the system is a hexagonal lattice of double-twist cylinders surrounded by negative disclinations~\cite{Nych2017,Duzgun2018,Metselaar2018}. These double-twist cylinders are structures with a topological skyrmion number $N = \frac{1}{4\pi}\int dx dy \vec{n} \cdot \left(\frac{\partial \vec{n}}{\partial x} \times \frac{\partial \vec{n}}{\partial y}\right) = \frac{1}{2}$, and they are therefore termed half-skyrmions (see Fig. \ref{fig:schematic}).

\begin{figure*}
\centering
\includegraphics[width=0.75\linewidth]{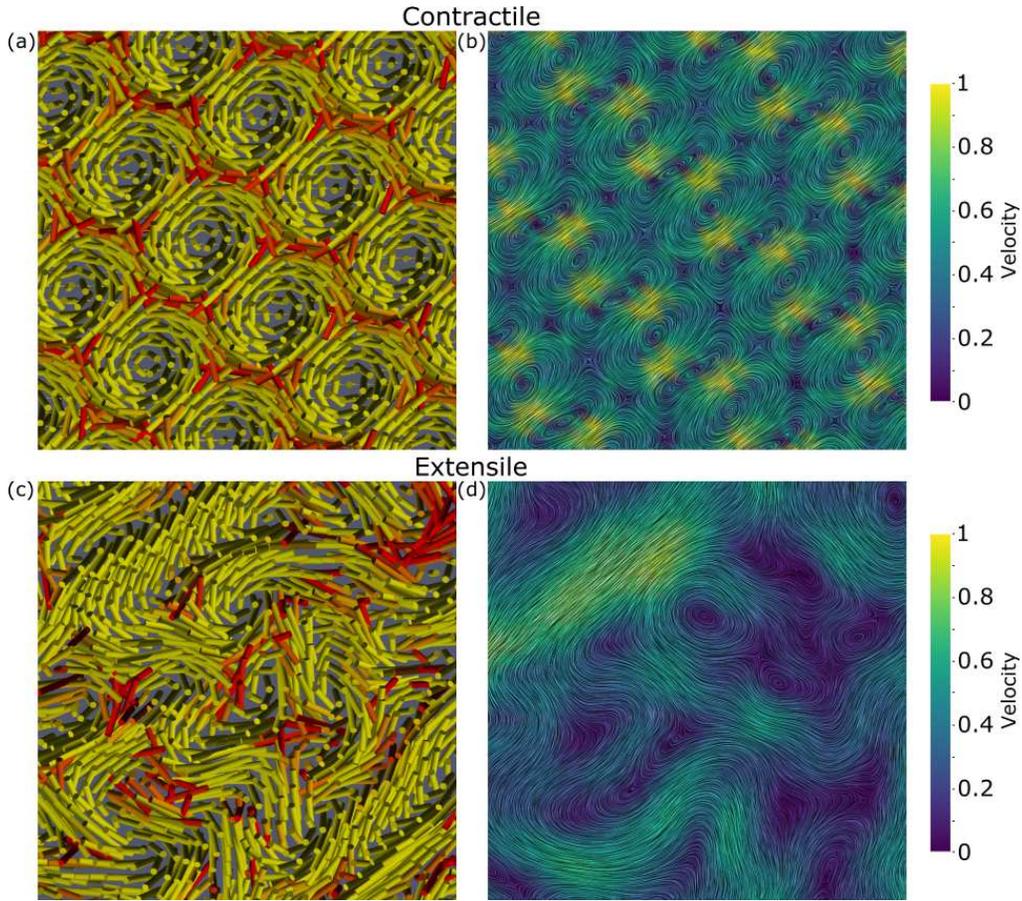}
\caption{An active two-dimensional blue phase with periodic boundary conditions. (a),(c) Director field, coloured by the magnitude of the order, ranging from disordered (red) to strongly ordered (yellow) for (a) contractile activity and (c) extensile activity. (b),(d) Velocity streamlines, coloured by the magnitude of the velocity normalised by its maximum value, ranging from slow (blue) to fast (yellow) for (b) contractile activity and (d) extensile activity. The contractile two-dimensional blue phase sets up a stable vortex lattice (a), (b), while the half-skyrmions cylinders in the extensile two-dimensional blue phase have dissociated, and the system is in an active turbulence regime (c), (d).}
\label{fig:example}
\end{figure*}

An example of the response to activity for a rod-like system with $\xi = 0.9$ can be viewed in Fig. \ref{fig:example} (see also Supplementary Movies 1 to 4). Upon application of contractile activity, the director field deforms from a double-twist cylinder configuration to a vortex-like `swirl' (Fig. \ref{fig:example}(a) and Supplementary Movie 1). The swirls all generate rotational flow with the same handedness, with regions of opposite vorticity in between (Fig. \ref{fig:example}(b) and Supplementary Movie 2). This type of vortex lattice is remarkably similar to those reported in active nematics with substrate friction \cite{Doostmohammadi2016}, where hydrodynamic screening stabilised the vortices, while here the additional length introduced by the intrinsic pitch of the cholesteric sets the vortex scale.

\begin{figure*}
\centering
\includegraphics[width=0.75\linewidth]{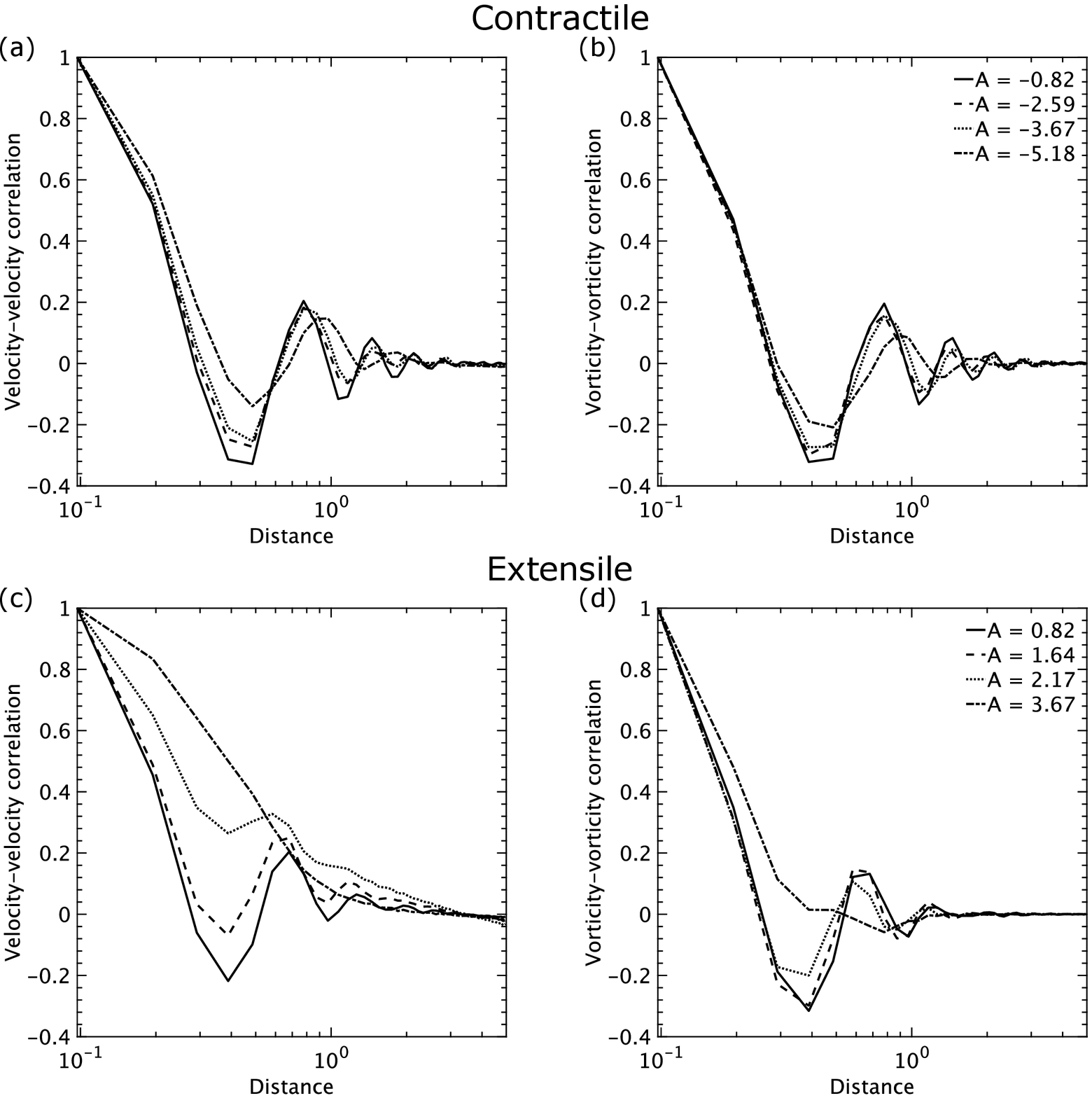}
\caption{Correlation functions for (a),(b) contractile and (c),(d) extensile activities. The distance has been scaled with respect to the characteristic pitch $p = 2\pi/q_0$. The dimensionless activity number $A = \pm\sqrt{\zeta/K{q_0}^2}$ characterises the ratio between the characteristic pitch and the active length scale $\sqrt{K/\zeta}$.}
\label{fig:correlations}
\end{figure*}

The hexagonal structure of the two-dimensional blue phase is stable for all contractile activity strengths considered. Fig.~\ref{fig:correlations}(a) shows the velocity-velocity correlation function $C_{uu}(r)=\langle\vec{u}(r,t)\cdot\vec{u}(0,t)\rangle / \langle \vec{u}(0,t)^2 \rangle$ for a range of contractile activities. The long-range order of the vortex lattice is clearly visible. The vorticity-vorticity correlation function $C_{\omega\omega}(r)=\langle\vec{\omega}(r,t)\cdot\vec{\omega}(0,t)\rangle / \langle \vec{\omega}(0,t)^2 \rangle$, where $\vec{\omega}$ is the vorticity, shows the same behaviour (Fig. \ref{fig:correlations}(b)). 

These two-dimensional blue phases are however unstable to extensile active stresses. An example of the director field (Fig. \ref{fig:example}(c) and Supplementary Movie 3) and the velocity field (Fig. \ref{fig:example}(d) and Supplementary Movie 4) show that under sufficiently large extensile activity the two-dimensional blue phase breaks up and active-turbulence-like behaviour emerges. This is further exemplified by the velocity-velocity correlation functions in Fig.~\ref{fig:correlations}(c), which show a reduction in long-range order with increasing extensile activity. Fig.~\ref{fig:correlations}(d) shows something unexpected, however: even though for intermediate extensile activities the correlation functions show that the long-range order is lost, the vorticity-vorticity correlation function still indicates the existence of a well-defined coherence length. This is because the active turbulence state is first developed for relatively coherent patches with the half-skyrmions moving as coherent units, in a turbulent-like fashion. Only for larger activity numbers is this coherence lost.

\begin{figure}
\centering
\includegraphics[width=\linewidth]{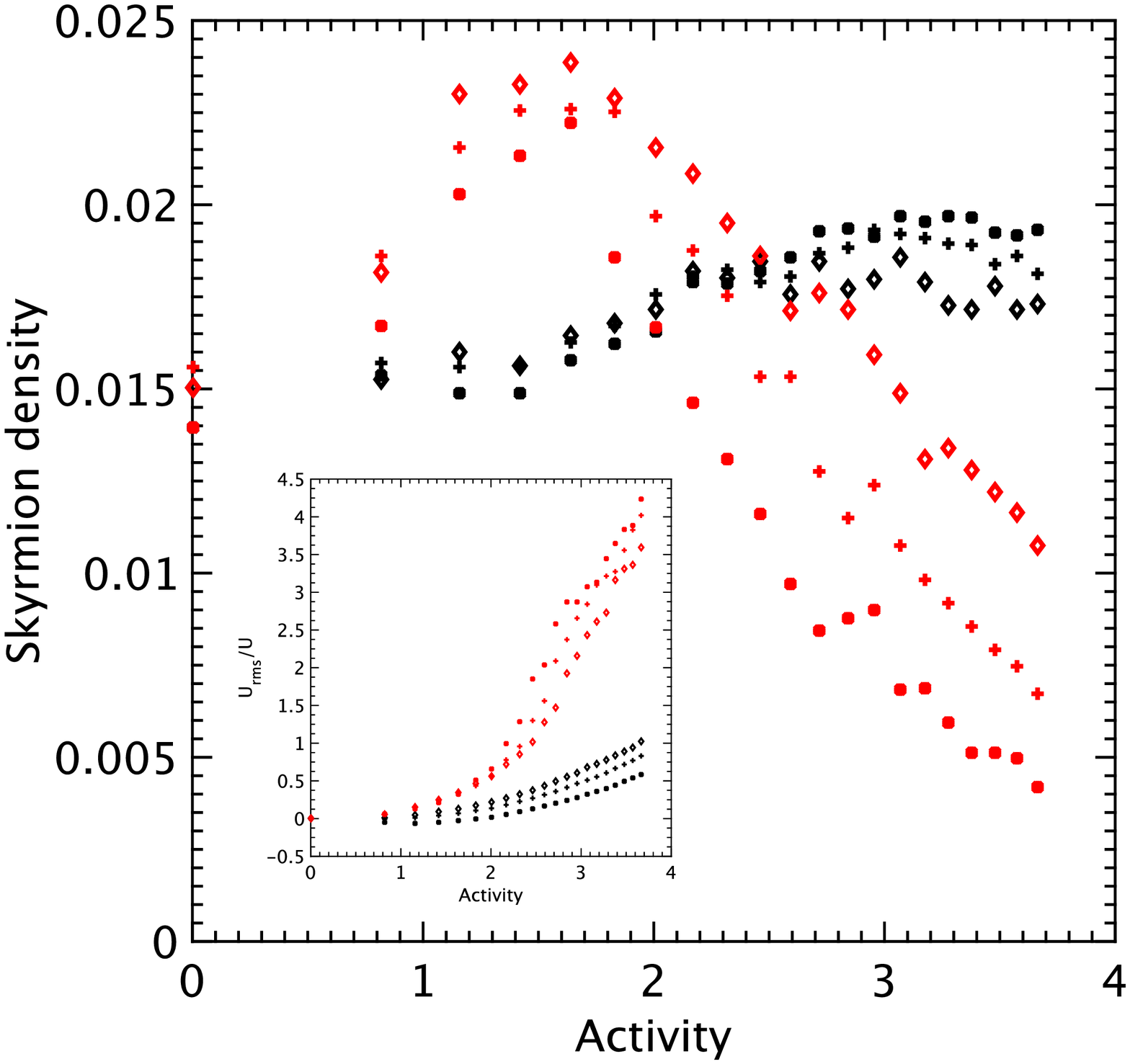}
\caption{The skyrmion density, defined as the number of skyrmions per unit area, averaged over time and over five different initial configurations. Black symbols are for contractile activity, and red symbols for extensile activity. The tumbling parameter $\xi$ is $0.5$ for diamonds, $0.7$ for plusses and $0.9$ for squares. Above a critical activity $A_\text{cr} \sim \sqrt{\zeta_\text{cr}/K{q_0}^2}$ the vortices dissociate in the extensile system and the number of skyrmions starts decreasing. As shown in the inset, the root mean squared velocity in the extensile systems increases faster above the critical activity. The root mean squared velocity is scaled by $U = K{q_0}^3/\eta$.}
\label{fig:defects}
\end{figure}

In addition, measurements of the number of half-skyrmions in the system highlight the difference between contractile and extensile driving (Fig.~\ref{fig:defects}). For contractile systems the number of half-skyrmions does not change substantially with increasing activity. The small increase in the number density is due to the stabilising effect of the contractile activity which helps the lattice become more regular so that it can fit slightly more half-skyrmions. For extensile systems, however, the behaviour is different. The initial increase in the number of half-skyrmions for extensile activity is due to the break up of short cholesteric stripes, which are present at low activity. More importantly, the number density of half-skyrmions drops significantly when the coherence is lost. The dissociation of half-skyrmions and onset of active turbulence occurs at approximately $\zeta_\text{cr} \approx 4K{q_0}^2$ ($A_\text{cr}\approx 2$) and it is accompanied by a faster increase in the root mean squared velocity $u_\text{rms}$ (Fig. \ref{fig:defects}; {\it inset}). This threshold can be explained by considering the competition between the pitch length set by the intrinsic chirality of the particles $p=2\pi/q_0$, and the active length scale set by the activity $\zeta$ and orientational elasticity $K$ of the system, $l_\text{a}\sim\sqrt{K/\zeta}$. As the activity is increased to the level where the active length scale becomes smaller than the pitch length, the half-skyrmions break up and the active turbulence is established. The number of half-skyrmions does not go to zero in the active turbulence regime, since there is no topological difference between a double-twist cylinder and a cholesteric stripe of finite length. In order to explain the different behaviour for contractile and extensile systems we next consider an isolated active half-skyrmion in an isotropic background.

\subsection{Isolated half-skyrmions}
\subsubsection{Theory}
We begin by presenting an analytical model of a single half-skyrmion in an isotropic fluid. Moving radially outwards, the director twists through an angle of $\pi/2$, from a vertical orientation at the centre of the skyrmion to a horizontal orientation in the azimuthal direction at the edge. The director field can be written as $\vec{n}(\vec{r}) = \left(\cos{\Phi(\phi)}\sin{\Theta(r)},\sin{\Phi(\phi)}\sin{\Theta(r)},\cos{\Theta(r)}\right)$, where the polar coordinates $\vec{r} = \left(r\cos{\phi},r\sin{\phi}\right)$ are introduced, and $\Phi(\phi) = \phi + \phi_0$ (with $\phi_0 = \pm \pi/2$ depending on handedness), and $\Theta(r) = r/R$ for $0 \leq r \leq R\pi/2$. 

We consider the three elastic modes, splay, twist and bend, usually written as the Frank free energy in terms of the director field $\vec{n}$,
\begin{equation}
\begin{split}
\mathcal{F}_\text{Frank} = \frac{K_1}{2}\left(\bnabla \cdot \vec{n}\right)^2 + \\ \frac{K_2}{2}\left(\vec{n} \cdot \bnabla \times \vec{n}\right)^2 + \frac{K_3}{2}\left((\vec{n} \cdot \bnabla)\vec{n}\right)^2.
\label{Frank}
\end{split}
\end{equation}
From equation \ref{Frank} it can be readily seen that a half-skyrmion is free of splay, but not of bend. It is well established that contractile stresses in active nematics drive an instability of the nematic state to splay deformations, while extensile stresses yield bend deformations \cite{Ramaswamy2010}. In chiral nematics however, the cholesteric order contributes a passive bend term that acts to screen the nematic splay mode. Instability only sets in when the activity exceeds a threshold, and only for extensile activity \cite{Whitfield2017}. Due to the cholesteric order in half-skyrmions, it can therefore be expected that a half-skyrmion is unstable to extensile stresses above a finite threshold in activity, but stable to contractile stresses. 

\begin{figure}
\centering
\includegraphics[width=\linewidth]{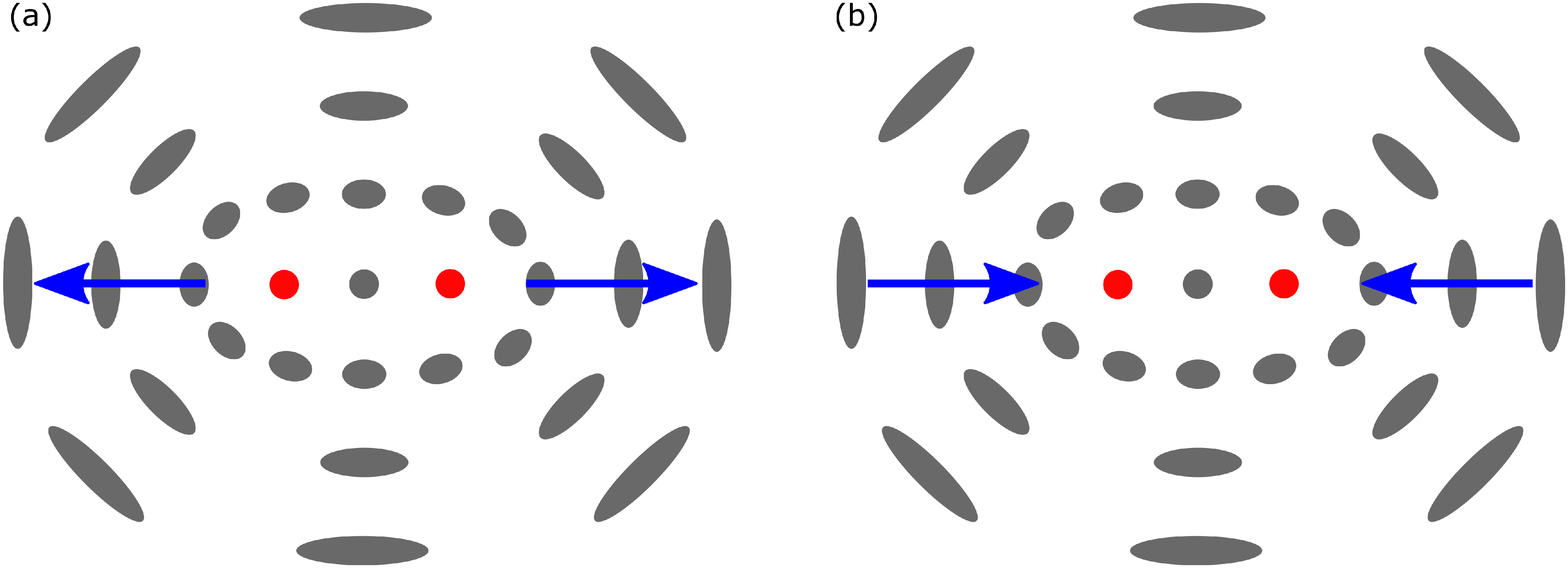}
\caption{Schematic of the half-skyrmion (with charge +1) split into two +1/2 topological defects. The approximate locations of the defects are indicated by red dots, and the blue arrows show the direction in which they self-propel. (a) Extensile active stresses lead to disclinations moving in the direction of their `heads', thus destabilising the half-skyrmion cylinder. (b) Contractile active stresses cause the disclinations to self-propel in the direction of their `tails', stabilising the half-skyrmion cylinder.}
\label{fig:schematic_defects}
\end{figure}

An intuitive way to understand the stabilising effect of contractile stresses observed in the simulations is by considering the half-skyrmion as a +1-topological defect that has escaped into the third dimension (Fig.~\ref{fig:schematic_defects}). If the +1 defect is split into two +1/2 topological defects, the direction of spontaneous motion of active +1/2-defects can be used to interpret the observed behaviour: extensile active stresses drive the two defects apart (Fig.~\ref{fig:schematic_defects}(a)), whereas contractile active stresses push the two defects together (Fig.~\ref{fig:schematic_defects}(b)). Since the undeformed half-skyrmion is a topological excitation without a singularity at its centre, introducing +1/2-defects will come with an energy cost. This explains the existence of a threshold for vortex dissociation in extensile systems.

It is obvious from a simple calculation that the undeformed half-skyrmion director field cannot generate rotational flow. If the Stokesian regime of equation \ref{eq:momentum} is considered, and the elastic terms in the stress tensor that generate backflow are neglected, the velocity obeys
\begin{equation}
\eta\Delta\vec{u} - \bnabla p = -\vec{f} = \zeta \bnabla \cdot \tens{Q} = \frac{3S\zeta}{2} \bnabla \cdot \vec{n}\vec{n},
\end{equation}
using the definition of $\tens{Q}$. The radial component of the force will be balanced by the pressure because of incompressibility, and the azimuthal gradient of the pressure is zero due to symmetry. The equation that has to be solved is therefore simply 
\begin{equation}
\eta\Delta u_\phi = -f_\phi = \frac{3S\zeta}{2rR}\sin{\frac{r}{R}}\left(r\cos{\frac{r}{R}} + R\sin{\frac{r}{R}}\right)\sin{2\phi_0}.
\end{equation} 
The symmetry of the undeformed director field, with $\phi_0= \pm \pi/2$, cannot give rise to a flow ($u_\phi = 0$). However, if the constant phase in $\Phi(\phi)$ is now perturbed, this will lead to a rotational flow.

The change in the constant phase is due to the director field deforming from pure bend deformations (in order to adapt the preferred twist) to a configuration that also accommodates splay to counterbalance the forces generated by the active stress. The case with constant phase $\phi_0=\pi/4$ deserves particular attention since it corresponds to the bend-splay +1 defect that has recently been studied in the context of living liquid crystals~\cite{Peng2016}. It was shown that a sufficiently high concentration of bacteria swimming through a predesigned pattern of a background liquid crystal gets attracted to a +1 bend-splay topological defect and forms a rotational flow around the core of the defect. Assuming boundary conditions $u_\phi(r=0) = 0$ and $u_\phi(r=R\sqrt{e}) = 0$, the velocity field of highly concentrated bacteria moving around the defect core was found to follow $u_\phi =  \frac{3S\zeta r}{16\eta}\left(-1 + 2 \log{\frac{r}{R}}\right)$ for a two-dimensional director field in the background liquid crystal, where there was no out-of-plane component. For the bend-splay half-skyrmion considered here, we find that the velocity is $u_\phi = \frac{3S\zeta r}{16\eta}\left(-1 + 2 \log{\frac{r}{R}} + 2\text{Ci}(2\sqrt{e}) - 2\text{Ci}(\frac{2r}{R})\right)$, with the cosine integral corrections due to the fact that the director is now varying in the out-of-plane dimension.

\subsubsection{Simulations}

\begin{figure}
\centering
\includegraphics[width=\linewidth]{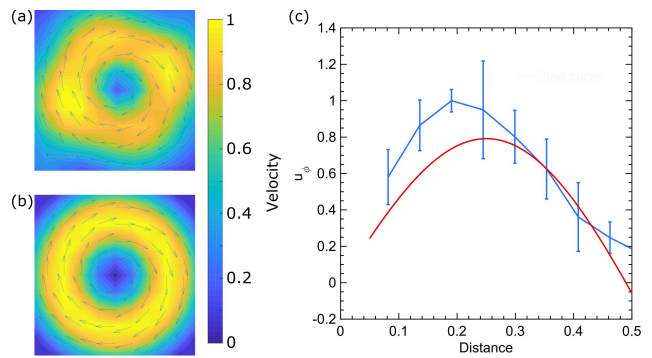}
\caption{The velocity field around an isolated active half-skyrmion from (a) simulations and (b) analytics. The arrows indicate the direction of the flow, with the vector length and the background colour indicating the magnitude normalised by the maximum velocity. (c) The radial average of the azimuthal velocity from the simulation (blue, with errorbars), and the velocity from the analytics (red). The distance is scaled by the characteristic pitch.}
\label{fig:single_skyrmion}
\end{figure}

The analytical result is only valid for single half-skyrmions and does not take into account the interactions between the half-skyrmions, nor the presence of the six -1/2-defects surrounding each cylinder in a blue phase. Therefore, in order to provide a closer comparison between our analytical prediction and the numerical results we next simulate half-skyrmions in an isotropic background by quenching from infinite temperature to $\gamma = 2.74$, with $q_0 = 2\pi/13$, and applying an external field in the out of plane direction to stabilise the half-skyrmion. To this end an external field free energy $\mathcal{F}_\text{ext} = -\vec{D} \cdot \tens{Q} \cdot \vec{D}$ is added to the total free energy, where $\vec{D}=0.15\hat{z}$ is the applied external field~\cite{Metselaar2018}. The resulting velocity field for a contractile activity $A = -3.27$ is shown in Fig.~\ref{fig:single_skyrmion}(a). In our analytical model we neglect the change in the magnitude of the nematic order parameter $\tens{Q}$ going from a half-skyrmion cylinder into the isotropic phase, but nevertheless obtain pleasing agreement with the simulation results (Fig.~\ref{fig:single_skyrmion}). It should be noted that the model only holds for contractile activity or small extensile activity. In case of large extensile activity the approximation that the director field is constant in time is no longer valid.

\subsection{Coherent rotation of a confined active blue phase}
\begin{figure}
\centering
\includegraphics[width=\linewidth]{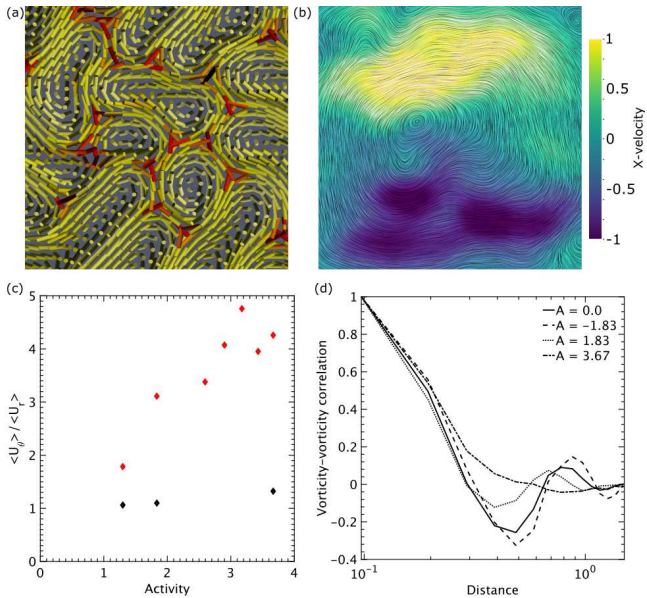}
\caption{Active two-dimensional blue phases in 30$\times$30 square confinement. Below the dissociation threshold extensile activity ($A = 1.83$) will leave half-skyrmions intact (a) and sets up a global rotational flow (b). The director field in (a) is coloured by the magnitude of the order, from disordered (red) to strongly ordered (yellow). The streamlines in (b) are coloured by the horizontal component of the velocity normalised by its maximum value, ranging from flowing to the left (blue) to flowing to the right (yellow). (c) The average orthoradial velocity divided by the average radial velocity shows that extensile activity (red diamonds) leads to coherent rotational motion, which is not present in the case of contractile activity (black diamonds). (d) The vorticity-vorticity correlation functions show that small extensile activity leaves the internal structure of the two-dimensional blue phase intact. The distance is scaled by the characteristic pitch.}
\label{fig:example_box}
\end{figure}

When active matter is confined, the otherwise turbulent-like motion can become coherent. For example cell monolayers~\cite{Doxzen13}, dense bacterial suspensions~\cite{Wioland13}, and microtubule/kinesin motor mixtures~\cite{Dogic18} can all self-organise into a single circulating unit inside a circular or square box. Therefore, we next ask if the active blue phase can behave in a similar way. Staying below the dissociation threshold in an extensile system, the half-skyrmions stay largely intact (see Fig.~\ref{fig:example_box}(a) and Supplementary Movie 5), and indeed move collectively to set up a coherent rotational flow (see Fig.~\ref{fig:example_box}(a),(b) and Supplementary Movies 5,6). The emergence of coherent rotation is best illustrated by calculating the ratio of the average orthoradial to radial velocity (Fig. \ref{fig:example_box}(c)). For extensile activity above the dissociation threshold circulation persists (Fig. \ref{fig:example_box}(c)), but without the hexagonal symmetry found in the two-dimensional blue phase. This is evident from the vorticity-vorticity correlation functions in Fig. \ref{fig:example_box}(d) ({\it dashed-dotted line}), which show that for large extensile activity the coherence is lost. For contractile activity no coherent rotational flow is observed as the blue phase remains unperturbed.

\section{Discussion}
Combining active stresses of self-propelled particles with two-dimensional helical ordering in the blue phase provides a framework for studying non-equilibrium topological states in active matter. Compared to the few existing theoretical works that treat chiral active materials \cite{Furthauer2012,Furthauer2013,Tjhung2017,Whitfield2017,Maitra2018}, we explicitly introduce topological states in chiral liquid crystals and connect the director structure to the resulting dynamics. We show that half-skyrmions are stable to contractile stresses, but that there is an active instability threshold for extensile active particles analogous to the pitch-splay mode in active cholesterics. This can be understood by regarding the topological excitation as a +1-defect that has escaped into the third dimension. A small perturbation will lead to two +1/2-defects, which are driven apart for extensile active stresses but driven together for contractile active stresses. For small extensile activity, the half-skyrmions do not dissociate, and they set up a coherent rotational flow when enclosed, reminiscent of collective cell behaviour in confinement. 

A number of directions for future work can be envisaged. For instance, the linear stability analysis performed for active nematics, smectics and cholesterics could be extended to the two-dimensional blue phase, introducing new symmetries to the framework. Secondly, the third dimension could be made finite, both in a linear stability analysis and in simulations, to study the onset of active instabilities along the axis of half-skyrmions. Thirdly, a three-dimensional active blue phase could be simulated to investigate whether these are also stable to contractile active stresses.

\begin{acknowledgments}
This project has received funding from the European Union's Horizon 2020 research and innovation programme under the DiStruc Marie Sk\l{}odowska-Curie grant agreement No. 641839. AD was supported by a Royal Commission for the Exhibition of 1851 Research Fellowship.
\end{acknowledgments}

\bibliography{chirality-1}

\end{document}